\documentclass[letters]{aa}  

\usepackage{graphicx}
\usepackage{txfonts}
\usepackage[colorlinks = true,
           linkcolor = blue,
            urlcolor  = blue,
            citecolor = blue,
            anchorcolor = blue]{hyperref}
\begin{document} 

    \titlerunning{ALMA 360 pc resolved dust temperature in a $z=6.9$ quasar}
   \title{ALMA 360 pc high-frequency observations reveal warm dust in the center of a $z=6.9$ quasar}

   \author{Romain A. Meyer
          \inst{1} \orcid{0000-0001-5492-4522}
          \and Fabian Walter \inst{2} \orcid{0000-0003-4793-7880}
          \and Fabio Di Mascia \inst{3}\orcid{0000-0002-9263-7900}
          \and Roberto Decarli \inst{4} \orcid{0000-0002-2662-8803}
          \and Marcel Neeleman \inst{5} \orcid{0000-0002-9838-8191}
          \and Bram Venemans \inst{6} \orcid{0000-0001-9024-8322}
          }

    \institute{Departement d'Astronomie, University of Geneva, Chemin Pegasi 51, 1290 Versoix, Switzerland\\
              \email{romain.meyer@unige.ch}
         \and
             Max-Planck Institute for Astronomy, K\"onigstuhl 17, 69118 Heidelberg, Germany
         \and 
         Scuola Normale Superiore, Piazza dei Cavalieri 7, 56126 Pisa, Italy
         \and
           Dipartimento di Fisica “G. Occhialini”, Università degli Studi di Milano-Bicocca, Piazza della Scienza 3, I-20126 Milano, Italy 
        \and National Radio Astronomy Observatory, 520 Edgemont Road, Charlottesville, VA, 22903, USA 
        \and Leiden Observatory, Leiden University, Niels Bohrweg 2, NL-2333 CA Leiden, The Netherlands }

   \date{Received --; accepted --}
 \abstract{
% 5 {} token are mandatory
 The temperature of the cold dust in $z>6$ galaxies is a potential tracer of Active Galactic Nucleus (AGN) and stellar feedback, and is the dominant source of uncertainty in inferring properties from the far-infrared (FIR) emission of these galaxies. We present the first resolved dust temperature map in a $z>6$ quasar host galaxy. We combine new 360 pc resolution ALMA Band 9 continuum observations with literature 190 pc Band 6 observations to derive the dust temperature and opacity at $0.1<r<0.5$ kpc scales in a $z=6.9$ luminous quasar host galaxy (J2348--3054). We find that the dust temperature (and opacity) increases at the center ($r<216$ pc) of the galaxy up to $T_d=73-88\ \rm{K}$, and potentially up to $T_d<149\ \rm{K}$ at $r<110$ pc. The combination of the resolved and integrated FIR Spectral Energy Distribution (SED) further reveal a dust temperature gradient and a significant contribution of the AGN hot dust torus at $\nu_{\rm{obs}}\gtrsim 700 \ \rm{GHz}$. By taking into account the torus contribution and resolved optically-thick emission, we derive a total infrared luminosity ($L_{TIR}=8.78\pm0.10)\times 10^{12}L_\odot$) and corresponding star-formation rate (SFR$=1307\pm15\ M_\odot\ \rm{yr}^{-1}$), that are at least a factor $\sim 3.6$ ($\sim0.57$ dex) lower than previous measurements assuming optically-thin emission. We compare the resolved dust temperature, mass and IR luminosity profiles to simulations where they are only reproduced by models in which the AGN radiation heats the dust in the center of the galaxy. Our observations provide evidence that dust in J2348--3054 cannot be assumed to be uniformly cold and optically thin. Whether J2348--3054 is representative of the larger population of high-redshift quasars and galaxies remains to be determined with dedicated high-resolution and high-frequency ALMA observations. }
  %\abstract

   

   \keywords{quasars: individual: J2348--3054 - galaxies: high-redshift - galaxies: ISM - galaxies: star formation}

   \maketitle
%
%-------------------------------------------------------------------

\section{Introduction}

The advent of ground-based sub-mm interferometry (ALMA/NOEMA) has opened a new window on the interstellar medium (ISM) of distant galaxies \citep[e.g.][ for a review]{Carilli2013}. Over the past two decades, fine structure lines and continuum emission in the rest-frame far-infrared (FIR) have been detected in increasing numbers of extragalactic sources. However, the interpretation of the FIR continuum emission relies on numerous assumptions about the underlying properties of cold ISM dust. The dust continuum modified blackbody emission peaks at rest-frame $\nu_{rest}\sim 3000-6000\ \rm{GHz}$, depending on the dust temperature. At high-redshift ($z\gtrsim 5$) high-frequency observations are thus required to accurately constrain the peak of the spectral energy distribution (SED) and infer the dust temperature, but have proved difficult to obtain. Even recent studies approach, but rarely go beyond, the peak of the dust SED at $z\gtrsim 6$ \citep[e.g.][]{Novak2019,Shao2022,Witstok2023, Algera2024,Tripodi2024}. FIR luminosities, star-formation rates (SFR), and sometimes gas mass estimates (when no suitable lines are present) are derived from the integrated dust SED, and are thus similarly uncertain. Even more challenging to obtain are resolved dust temperature measurements (and thus resolved SFR, gas and dust mass profiles), which require both high-resolution and at multiple frequencies, including at high frequencies. 
Resolved dust temperature measurements are however crucial to investigate the assumptions made about dust properties when analysing unresolved data, and offer a new route to probe the interstellar medium (ISM) of distant galaxies.

Of particular interest is the interaction between the active galactic nucleus (AGN) and the surrounding ISM. Numerous studies have searched for evidence of AGN feedback in the host galaxies of luminous $z>6$ quasars powered by supermassive black hole with masses ($M_{BH}>10^8-10^9 \ M_\odot$) and accretion rates close to the Eddington accretion rate, and thus expected to be the best places to look for AGN feedback in the first billion years. The majority of studies have focused on [\ion{C}{II}] outflows, with mixed results \citep{Maiolino2012,Cicone2015, Bischetti2019,Novak2020,Izumi2021,Tripodi2022,Shao2022,Meyer2022,Tripodi2024}. 
However, other forms of feedback signatures predicted by theoretical models, such as highly-excited gas tracers \citep[e.g.][]{Carniani2019} or warm ($\gtrsim 100\ \rm{K}$) dust, have been comparatively less explored. In a recent study, \citet{DiMascia2021} show that radiation from the AGN is expected to heat the ISM in its vicinity. Although the mass of warm dust heated in this way is negligible, it can significantly affect the integrated dust SED at longer wavelengths. Disentangling the effect of the AGN on the dust is not only an important new tracer of AGN feedback, but is also crucial to measure accurately the star-formation heated dust continuum emission  of their host galaxies to characterise them \citep[e.g.][]{Scoville1976, Casey2012,Tsukui2023,FernandezAranda2025}. So far however, such measurements have been impossible at $z>5$ due to the difficulty in obtaining the necessary high-resolution and high-frequency observations. Currently, resolved (i.e., sub-kpc) observations of high-redshift ($z>5$) sources are still rare and limited to a single continuum frequency around the [\ion{C}{ii}] line \citep[e.g.][]{Venemans2019, Walter2022, Meyer2023, Meyer2025,Neeleman2023,Rowland2024}. 

J2348--3054 is a luminous broad absorption line (BAL) quasar at $z=6.9$ discovered in VIKING imaging data \citep{Venemans2013}. It hosts a black hole with a mass of $0.6-2.0\times10^{9}\ M_\odot$ \citep{DeRosa2014, MAzzucchelli2017, Farina2022}. The host galaxy was first detected in unresolved [\ion{C}{II}] and ALMA Band 3 continuum \citep{Venemans2015, Venemans2016, Venemans2020}, and more recently was observed at $190\ \rm{pc}$ resolution in [\ion{C}{II}] and continuum by \citet{Walter2022}. The galaxy is infrared- and [\ion{C}{II}]-luminous ($L_{\rm{TIR},8-1000\mu\rm{m}}=3.2\times10^{13}\ L_\odot,\ L_{[\ion{C}{II}]}=1.9\times10^9\ L_\odot$), massive (dynamical mass $M_{\rm dyn}=10^{10.3\pm0.3}\ M_\odot$), with a very compact host galaxy (the central $r=530\ \rm{pc}$ containing $\sim 90\%$ of the continuum flux) and shows kinematics compatible with a rotating disk. Based on the surface brightness observed in the central resolution element, \citet{Walter2022} suggested that the continuum dust emission is optically-thick, with a dust temperature $T_d>130\ \rm{K}$ associated with an Eddington-limited starburst ($\Sigma_{SFR}> 10^4 M_\odot\ \rm{yr}^{-1} \ \rm{kpc}^{-2}$) in the central hundred parsec. 

In this Letter, we present resolved ALMA Band 9 continuum observations (rest-frame 4623 GHz) probing the peak of the dust SED of J2348-3054. Combined with the previous Band 6 190 pc resolution data \citep{Walter2022}, these observations enable us to derive the resolved dust temperature profile and dust opacity for the first time at this redshift. We confirm the presence of an increase in the dust temperature in the central hundred parsec and discuss the role of the quasar in heating the dust in the central region. This Letter showcases how ALMA is able to resolve the ISM properties of the first galaxies in the Universe. Throughout this Letter we use a concordance cosmology with $\Omega_\Lambda = 0.7, \Omega_M = 0.3, H_0 = 70\ \rm{km\ s}^{-1}$.

\section{The resolved dust properties of a $z=6.9$ quasar}

\begin{figure*}
    \centering    \includegraphics[width=0.9\linewidth]{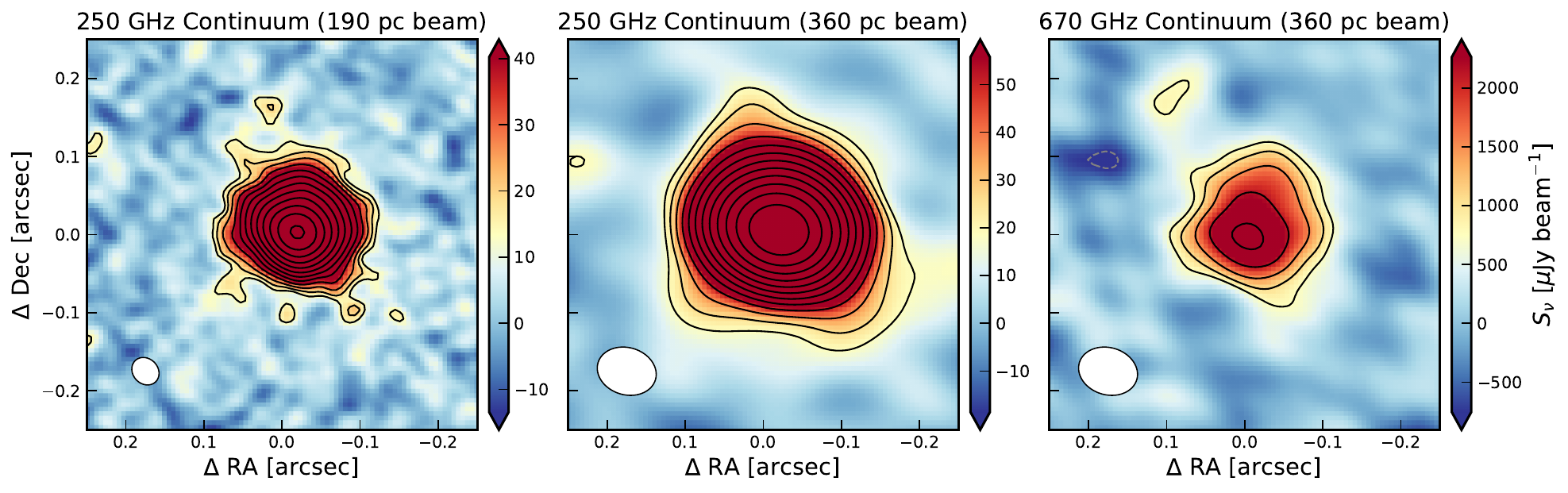} 
    \caption{Resolved continuum measurement of J2348--3054. \textbf{Left: } Rest-frame 250 GHz continuum map reproduced form \citet{Walter2022}. \textbf{Middle: } 250 GHz continuum map, convolved to the new Band 9 beam. \textbf{Right: } New Band 9 continuum map at rest-frame  $670 \ \rm{GHz}$. In the three panels, the solid (dashed) contours show the positive (negative) contours at $+(-)3\sigma$ levels, increasing in power of $\sqrt{2}$. The beam is shown in the lower left corner in white.  }
    \label{fig:fig1}
\end{figure*}
We observed the $z=6.9$ quasar J2348--3054 in configuration C43-6 in ALMA Band 9 at a frequency of $670\ \rm{GHz}$ (Program ID \#2021.1.01350.S, PI: Meyer). The final on-source time is $48\ \rm{min}$. J1924-2914 was used a bandpass calibrator, and J2349-3133 as phase calibrator. The data was reduced directly by the ALMA ARC Node. Baselines exceeding a phase RMS of $>100$ deg were manually flagged to minimize decoherence. Importantly, the maximum recoverable scale ($0\farcs9$) is similar to that of the $250\ \rm{GHz}$ continuum observations \citep{Walter2022}. We imaged the Band 9 continuum using \texttt{CASA} version 6.2.1-7 \citep{THECASATEAM2022} with multiscale cleaning down to $2\sigma$ and a robust Briggs weighting of $r=0.5$  using the entire bandwidth ($15\ \rm{GHz}$) with a pixel scale of $0\farcs005$. Visual inspection of the datacube did not reveal significant absorption or emission in the spectrum. The final $670\ \rm{GHz}$ continuum map has a synthesized beam of $0\farcs076\times0\farcs06$, corresponding to a FWHM of $\sim 360\ \rm{pc}$ and an effective radius of $r=216\ \rm{pc}$, corresponding to a circular area with equivalent area to the beam (for comparison, the beam of the Band 6 data has an effective radius $r=110\ \rm{pc}$). The rms of the continuum map is 215 $\mu\rm{Jy\ beam}^{-1}$. Finally, we convolve the $250\ \rm{GHz}$ continuum map from \citet{Walter2022} down to the $670\ \rm{GHz}$ continuum beam using the \texttt{SpectralCube} package\footnote{\url{https://spectral-cube.readthedocs.io/en/latest/}}, resulting in an rms of $6.2\ \mu\rm{Jy\ beam}^{-1}$ at $250\ \rm{GHz}$.

We find a peak continuum flux density of $3.2\pm0.2 \ \rm{mJy}$ at a position of RA=23:48:33.346, Dec=-30:54:10.307 (ICRS). We then integrate the $670\ \rm{GHz}$ continuum flux over an aperture of $r=0\farcs2$ centered on this peak pixel, and find $S_{\nu=670\ \rm{GHz}}= 9.1\pm1.1\ \rm{mJy}$ (where we have applied residual-scaling \citep[see e.g.][]{Jorsater1995,Walter1999} as implemented in the \texttt{interferopy} Python package). The aperture radius of $r=0\farcs15$ is chosen with a curve-of-growth analysis, where we find that the flux is consistent within uncertainties for all $r>0\farcs15$ apertures.

We show the resolved $670\ \rm{GHz}$ continuum observations alongside the original and convolved Band 6 continuum data in Fig. \ref{fig:fig1}.  The contours of the $250\ \rm{GHz}$ continuum are more tightly spaced than that of the $670\ \rm{GHz}$ continuum, suggesting a radial evolution of $S_{\nu=250\ \rm{GHz}}/S_{\nu=670\ \rm{GHz}}$ and thus the dust emission properties. To quantify this, we fit the SED in each pixel using the Band 6 and Band 9 resolved data (we further show in Appendix \ref{app:dirtybeam} that our results are not affected by the imprint of the two different dirty beams in the residuals). As \citet{Walter2022} showed, the dust emission is likely optically-thick in the inner center of the quasar host galaxy. We therefore fit the dust SED using the dust emission equation in the optically-thick regime and apply Cosmic Microwave Background contrast and heating corrections as per \citet{DaCunha2015}
\begin{equation}
S_\nu\,=\,\Omega_A\,[B_\nu(T_d)-B_\nu(T_{\rm CMB})]\,[1-\exp(-\tau_\nu)]\,(1+z)^{-3},
\label{eq:snu}
\end{equation}
where $S_\nu$ is the frequency--dependent dust continuum flux density, $\Omega_A$ is the solid angle corresponding to the chosen aperture or beam size, $B_\nu(T_d)$ and $B_\nu$($T_{\rm CMB}$) are the black body emission ($B_\nu(T)=2\,h\,\nu^3\,c^{-2} [\exp(h\,\nu/(k_{\rm b} T)) -1]^{-1}$) from the dust and cosmic microwave background (CMB), with $h$ the Planck constant, $c$ the speed of light, $k_B$ the Boltzmann constant, and $\tau_\nu$ is the optical depth (as a function of frequency $\nu$) of the dust. In this work, we adopt the prescription of \citet{Draine2003} $\tau_\nu=\kappa_0 (\nu / \nu_{\rm ref})^\beta\,M_{\rm dust}\,A^{-1}$, where $\kappa_0=13.9$\,cm$^2$\,g$^{-1}$ is the absorption coefficient per unit dust mass, $\beta$ is the emissivity index, $\nu_{\rm ref}=2141$\,GHz is the reference frequency for the dust emissivity $M_{\rm dust}$ is the dust mass and $A=\pi r^2$ is the area of the emitting region. With the dust opacity specified in this way, we can fit the resolved dust SED for the dust temperature, mass and $\beta$ slope. In this work, we assume $\beta=1.8$ which corresponds roughly to the median value found in statistical samples of galaxies and quasars observed in multi-band FIR observations at $z>5$ \citep[e.g.][]{Witstok2023,Tripodi2024}. Consequently, the only two free parameters of our fit are the dust temperature and mass (see Appendix \ref{app:thin_thick_differences} for a discussion of the non-degenerate nature of our temperature constraints).

\begin{figure*}
    \centering
    \includegraphics[width=0.76\linewidth]{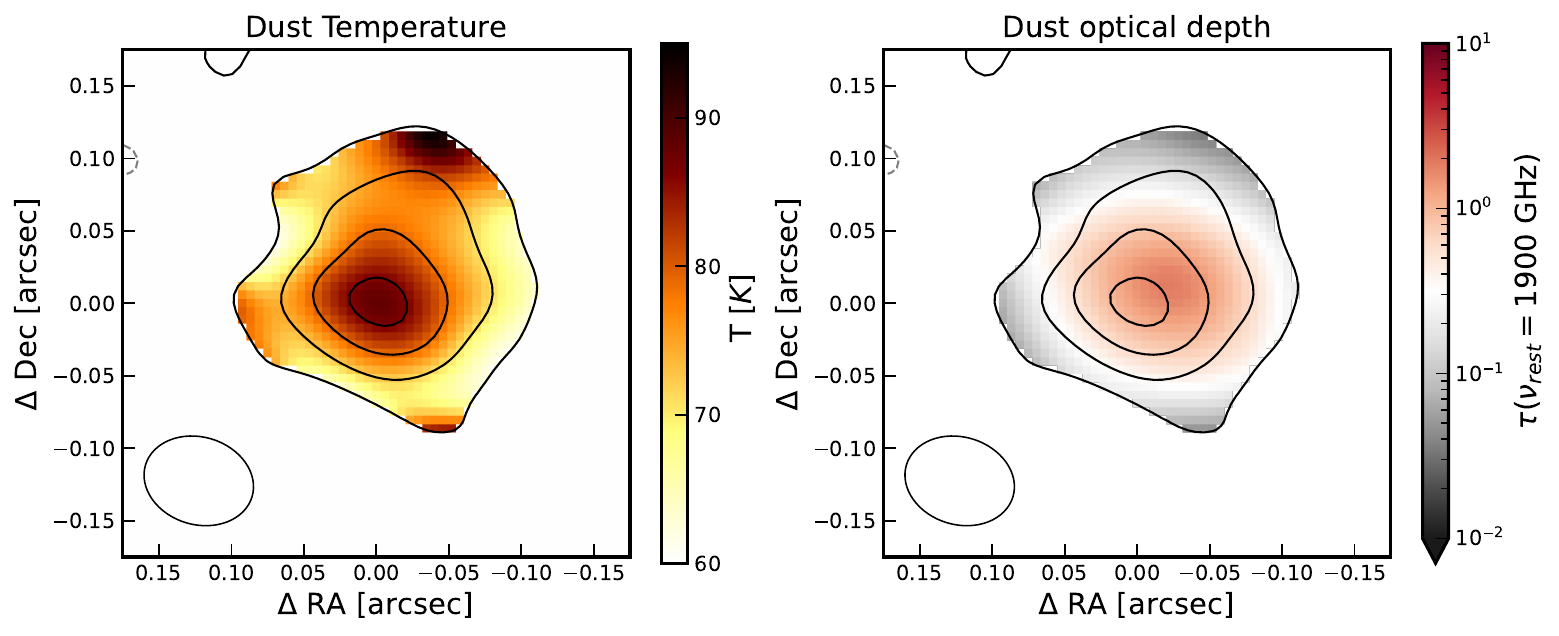}
    \caption{Maps of the dust temperature (top left) and dust optical depth (top right), obtained from our best-fit optically-thick dust SED models to the $360\ \rm{pc}$ resolution Band 6 and Band 9 continuum observations. We only fit pixels where the continuum is detected at SNR$>3$ in both bands. The median uncertainty in the dust temperature is $\Delta T =4.4\ \rm{K}$ and $\Delta\tau = 0.06$ for the dust opacity. Our observations reveal an incresease of dust temperature and optical depth at the center of J2348--3054.}
    \label{fig:fig2_maps}
\end{figure*}

We show the resolved dust temperature and optical depth in Fig. \ref{fig:fig2_maps}. We only fit pixels where the $670\ \rm{GHz}$ and $250\ \rm{GHz}$ continuum is detected at SNR$ >3$, which essentially reduces to the $670\ \rm{GHz}$ SNR$ >3$ criterion given the deeper sensitivity of the $250\ \rm{GHz}$ observations. The dust temperature peaks in the central beam to a beam-averaged temperature $88\pm2\ \rm{K}$, with an optical depth of $\tau_d= 1.8$ at $\nu_{rest}=1900\ \rm{GHz}$. Accordingly, the  central ($r<216\ \rm{pc}$) total infrared luminosity ($8-1000\ \mu\rm{m}$) is $L_{\rm{TIR}} =  4.8\pm0.5 \times 10^{12}\ L_\odot$. Assuming that dust is heated by star-formation we would derive a central SFR of $717\pm75 \ M_\odot\ \rm{yr}^{-1}$ using the \citet{Kennicutt2012} relation. The corresponding star-formation rate densities (SFRD) $\Sigma_{SFR}=4.9\times10^{3}\ M_\odot \ \rm{yr}^{-1}\ \rm{kpc}^{-2}$. These values could be lower if the central pixel is contaminated by AGN torus hot dust emission and/or the cold dust is in part heated by the AGN, two possibilities we will discuss in Section \ref{sec:integrated_sed} and \ref{sec:feedback}, respectively.

We note that the lower resolution of our observations \citep[$\sim$ twice that of][]{Walter2022} results in lower dust temperatures, SFR, dust masses and opacities in the central resolution element as the radial density decreases rapidly away from the center of the galaxy. We can still treat the observed continuum in the central beam $360\ \rm{pc}$ resolution as an upper limit to the flux  at $190\ \rm{pc}$ resolution \citep{Walter2022}. Following the same method as above, we find an upper limit on the a central ($r<110\ \rm{pc}$) dust temperature $T_d < 149\pm5\ \rm{K}$, a total infrared luminosity ($8-1000\ \mu\rm{m}$), $L_{\rm{TIR}}< (10.5\pm0.6) \times 10^{12}\ L_\odot$, a central  SFR of $< 1567\ M_\odot\ \rm{yr}^{-1}$ and SFRD $\Sigma_{SFR}< 4.1\times10^{4}\ M_\odot \ \rm{yr}^{-1}\ \rm{kpc}^{-2}$, in agreement with \citet{Walter2022}.

\section{Constraining cold/warm ISM dust and the AGN hot dust torus emission}
\label{sec:integrated_sed}
\begin{figure}
    \centering
    \includegraphics[width=\linewidth]{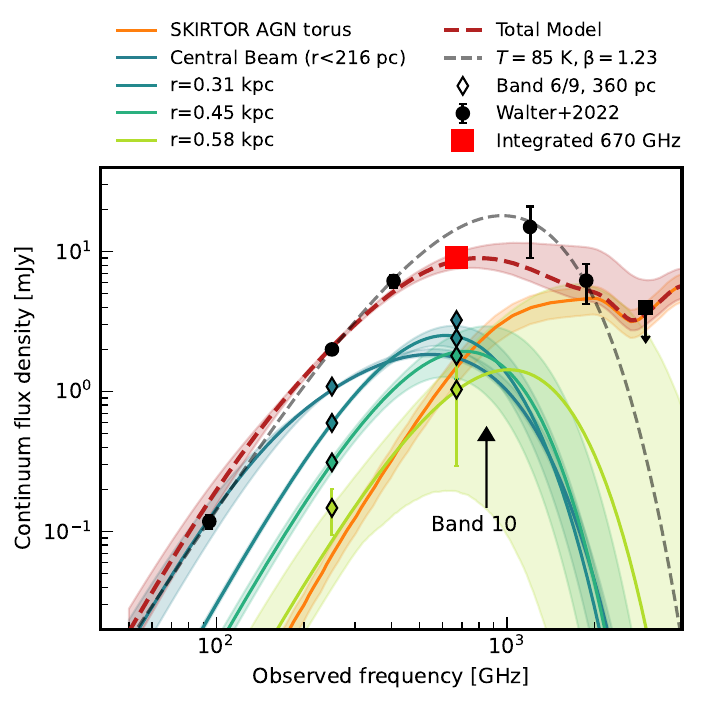}
    \caption{FIR SED of J2348--3054. We show the constraints from ALMA, ACA and Herschel/PACS \citep{Venemans2017,Walter2022} in black as well as our Band 9 integrated measurement in red. The dashed grey line show the single optically-thin emission model of \citep{Walter2022} which does not reproduce the Band 9 continuum. The 360 pc resolved 250/670 GHz continuum fluxes are present in colored diamonds. The total SED, composed of the sum of the individual annuli SED and the SKIRTOR torus (shaded lines), is shown in dashed red and shaded area. Further ALMA Band 10 (resolved) observations could constrain the properties of the dust torus which would contribute $\sim 2.7\ \rm{mJy}$ in the central pixel at $\nu\sim 875\ \rm{GHz}$. }
    \label{fig:total_sed}
\end{figure}

An important consequence of our results concerns the standard interpretation of the FIR SED with a single, cold dust SED greybody model. As shown in Figure \ref{fig:total_sed}, the ALMA Band 9 continuum data excludes the optically-thin model fitted by \citet{Walter2022} with a single temperature of $85\ \rm{K}$. 
The summed $SNR>3$ Band 6 and Band 9 pixels (see previous section) accounts for $\sim 80\%$ of the total flux, as cold ($T_d\lesssim 60\ \rm{K}$) dust emission is likely missing from the outskirts of the galaxy. Incidentally, the observed SED implies that the Herschel/PACS fluxes at $\nu_{\rm{obs}}\gtrsim 1000\ \rm{GHz}$ contain a non-negligible contribution from the AGN hot dust torus emission.

We thus proceed to fit simultaneously the resolved continuum at $250$ and $670$ GHz and the integrated FIR emission using all literature data at $100-2000$ GHz. In order to do this, we divide the resolved continuum data in four annuli starting with the central resolution element ($r<216$) up to $r=0\farcs15$. We then fit the dust continuum in each annuli (assuming optically-thick emission), and additionally include the contribution from hot dust torus emission in the central beam. To model the AGN torus we use \texttt{SKIRTOR} AGN models  \citep{Stalevski2012,Stalevski2016} with a nearly face-on geometry ($i=10$\textdegree), motivated by the fact that J2348--3054 is a luminous type 1 quasar and the face-on disk [\ion{C}{II}] kinematics \citep{Walter2022}. We define the total log-likelihood as the sum of the individual log-likelihoods in each annuli and that of the integrated SED (see further Appendix \ref{app:dust_fit}).

We show the resulting best fits and uncertainties in Fig. \ref{fig:total_sed}, and refer the reader to Appendix \ref{app:dust_fit} for details of the dust properties in each annuli and the parameters posterior distribution \footnote{For future reference the best-fit SKIRTOR AGN model is the \textit{t3\_p1.5\_q1.5\_oa80\_R10\_Mcl0.97\_i0\_sed.dat} model. We refer to \citet{Stalevski2016} for the naming convention and physical parameters of the model.}. We find that the central temperature decreases to $T_d(r<216\ \rm{pc})=72^{+2}_{-1}$~K when including an AGN hot torus contribution, although the strength of the torus emission and the central dust temperature are highly anti-correlated. Further high-resolution observations in Band 10 could break this degeneracy, where the AGN torus emission should dominate the central beam with a flux of $f_\nu(875\ \rm{GHz})=2.7\ \rm{mJy}$. Nonetheless, we still find lower dust temperatures at larger radii $T_d\sim60-66\rm{K}$ (see Appendix \ref{app:dust_fit}), indicating that the dust temperature gradient discussed in the previous section is robust. 

Finally, we find that the summed FIR luminosity of the four annuli is $L_{FIR} = (8.78\pm0.10)\times 10^{12}L_\odot$, corresponding to SFR$=1307\pm15\ M_\odot\ \rm{yr}^{-1}$. This is $3.6\times$ lower than the previously-reported value of \citet{Walter2022} based on a single, optically-thin emission model, highlighting the importance of high-resolution observations to unveil optically-thick dust emission and potential AGN contamination. 

\begin{figure*}
    \centering
    \includegraphics[width=\linewidth]{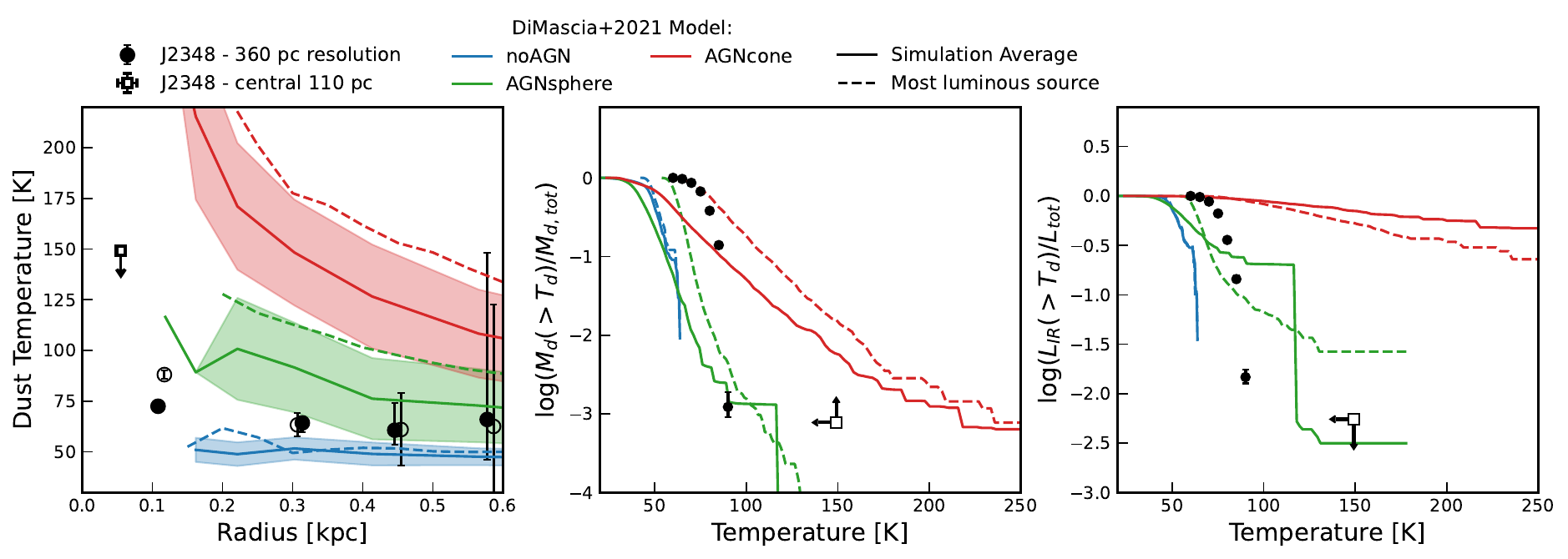} 
    \caption{Comparison between our observations and \citet{DiMascia2021} simulation. We show the simulated profiles for the most luminous source in the simulations in dashed lines, and the average profile of all sources in the simulation with full lines. We show the constraints from the matched-resolution (360 pc) ALMA Band 9 and Band 6 data in full datapoints, and the central limits on the inner $r<110\ \rm{pc}$ region with an open square (see further text). \textbf{Left:} Radially-averaged dust temperature profile. We indicate the dust temperature inferred from the simultaneous fit in full circles, and that from independent fit not taking the total FIR SED into account with open circles (see Appendix \ref{app:dust_fit}).  \textbf{Middle:} Mass fraction as a function of the dust temperature.  \textbf{Right:} Total infrared luminosity (8-1000 $\mu\rm{m}$) fraction as a function of the dust temperature. }
    \label{fig:simulation_comparison}
\end{figure*}

\section{Is AGN radiation heating the dust in the center of J2348--3054?}
\label{sec:feedback}

Our results show that the dust temperature in J2348-3054 is $\sim 60\ \rm{K}$ in the outskirts ($\sim 0.3-0.5\ \rm{kpc}$) and rises up to $70-90\ \rm{K}$ in the central $r<360\ \rm{pc}$ (Fig. \ref{fig:fig2_maps}), and even up to $149\pm5\ \rm{K}$ in the central $r<110\ \rm{pc}$. An interesting question is whether the luminous BAL quasar or an associated intense starburst is primarily responsible for heating the dust. We compare our observations to three different simulations from \cite{DiMascia2021}: 1) no AGN feedback, 2) spherical AGN feedback, and 3) conical AGN feedback (Fig. \ref{fig:simulation_comparison}). In particular, we compare the dust temperature profile, dust mass and luminosity fraction as a function of temperature for the most luminous source in their simulations ($M_{1450} = -21.4, -21.3, -25.4$ in their \textit{noAGN}, \textit{AGNsphere}, \textit{AGNcone} runs). We find that the simulations without AGN feedback do not reproduce the warm dust temperature profile observed in the center of J2348--3054, and our observations are in general agreement with the \textit{AGNsphere} model. 

We have shown previously that the FIR SED of J2348--3054 indicates the presence of an AGN hot dust torus component at $\nu_{\rm{obs}}\gtrsim 1000\ \rm{GHz}$ (Fig. \ref{fig:total_sed}), whilst the resolved continuum observations at 250 and 670 GHz provides evidence for warm  dust in the core of the galaxy. Although we cannot formally exclude that the central FIR luminosity and dust temperature in J2348-3054 is due to a starburst, these results and the better agreement of our resolved temperature observations with a model including AGN feedback (Fig. \ref{fig:simulation_comparison}) suggest that the high dust temperature is most likely due to AGN radiation heating the dust.

An important consequence of AGN radiation heating the dust is that the previous estimates of both the central and total SFR are too high. As a first-order estimate of this overestimation, we can consider excluding the contribution from the central $r<216\ \rm{pc}$ region, which accounts for $35\%$ of the continuum emission at $670\ \rm{GHz}$. Since the high temperature is likely AGN-related and not due to star-formation, the central FIR luminosity should not be converted to SFR. We note that our analysis in Section \ref{sec:integrated_sed} has revised the FIR luminosity by a factor $3.6$ compared to a single optically-thin greybody estimation \citep{Walter2022}, implying a total $\sim 5\times$ overestimation of the total SFR when excluding the central resolution element.
We note that this is a very rough estimate: in reality, the AGN can heat dust beyond the central beam, and conversely, the central beam is not completely devoid of star-formation. A more quantitative (but model-dependent) estimate of the overestimation of the SFR can be established with simulations, with \citet{DiMascia2023} suggesting that the SFR in J2348--3054 might be overestimated by a factor $\sim 10$, significantly above our back-of-the-envelope calculation. If confirmed in this and other high-redshift quasars, this would alter our view of $z>6$ quasar host galaxies as extreme starbursts, and bring their SFR in line with those observed in non-active galaxies such as REBELS and ALPINE \citep{Bouwens2020, Schouws2022, Bethermin2020,Schaerer2020}.

A number of caveats apply to our analysis. Firstly, the luminosity of the most-luminous source in the \textit{AGNsphere} model is much lower than that observed for J2348--3054. However, for the \textit{AGNcone} simulation, the temperature profiles are relatively similar between the most luminous source (as luminous as J2348--3054 this time), and the average profile of all objects in the box. Secondly, whilst the simulations match best the observations when AGN feedback is switched on, other models and simulations might reproduce the observed temperature profiles. Finally, J2348--3054 is amongst the most compact and FIR-luminous high-redshift quasars observed with ALMA so far, which could either support the AGN heating scenario detailed here, or simply be an outlier in terms of SRFD.

\section{Summary and future prospects}

We have reported the first high-frequency ($>600\ \rm{GHz}$) and high-resolution ($360\ \rm{pc}$) observations of a $z\sim 7$ quasar host galaxy, resolving the dust temperature and optical depth across its host galaxy. These observations sample the peak of the dust SED and spatially resolve it at 360 parsec resolution, leading to the following important results:

\begin{itemize}
    \item We report the presence of a warm ($88\pm2\ \ \rm{K}$), optically-thick ($\tau_{\nu,\rm{rest}=1900\ \rm{GHz}}\sim 2$) dust component at the center ($r<216\ \rm{pc}$) of J2348--3054. Including a contribution from the AGN hot dust torus, we find a slightly reduced dust temperature $T_d= 72_{-1}^{+2}\ \ \rm{K}$, although further ALMA Band 10 high-resolution observations are necessary to conclude on the exact AGN torus contribution.
    \item The approximation of the unresolved dust emission as optically-thin is inadequate, leading to overestimated measurements of the FIR luminosity. Consequently, we show that the SFR of J2348--3054 is likely overestimated by a factor $\sim 3.6$. This could rise to a factor $\sim 5$ ($\sim0.7$ dex) if the contribution of the central $216\ \rm{pc}$ parsec is removed as the dust in this region is likely heated by AGN radiation.
    \item Comparison to simulations by \cite{DiMascia2021} show that the dust is likely heated by AGN radiation, with a strong preference for their spherical feedback prescription. 
\end{itemize}
ALMA high-frequency, resolved observations of high-redshift sources have been missing until now. However, these results, obtained in only 48 minutes, show the potential of such observations to reveal and resolve the ISM properties of the most distant IR-luminous sources. With an abundance of lower-frequency high-resolution observations in the archive, it appears that a minimal investment in ALMA Band 8 to 10 observations would yield considerable progress in the study of galaxies and quasars in the first billion years of the Universe.

\begin{acknowledgements}
We thank the anonymous referee for an insightful report that improved this Letter.
RAM acknowledges support from the Swiss National Science Foundation (SNSF) through project grant 200020\_207349. FW acknowledges support from the ERC Advanced Grant 740246 (Cosmic\_Gas). 
This paper makes use of the following ALMA data: ASA/JAO.ALMA\#2018.1.00012.S, ASA/JAO.ALMA\#2021.1.01350.S.  ALMA is a partnership of ESO (representing its member states), NSF (USA) and NINS (Japan), together with NRC (Canada), NSTC and ASIAA (Taiwan), and KASI (Republic of Korea), in cooperation with the Republic of Chile. The Joint ALMA Observatory is operated by ESO, AUI/NRAO and NAOJ.
This work made use of the following Python packages: \emph{numpy} \citep{Numpy2020}, \emph{matplotlib} \citep{Hunter2007}, \emph{scipy} \citep{Virtanen2020}, \emph{Astropy} \citep{TheAstropyCollaboration2018,TheAstropyCollaboration2022}, \emph{interferopy} \citep*{interferopy}.
\end{acknowledgements}

\bibliographystyle{aa}
\bibliography{bib}

\appendix

\section{Impact of the dirty beam and residuals}
\label{app:dirtybeam}

\begin{figure*}
    \centering
    \includegraphics[width=0.8\linewidth]{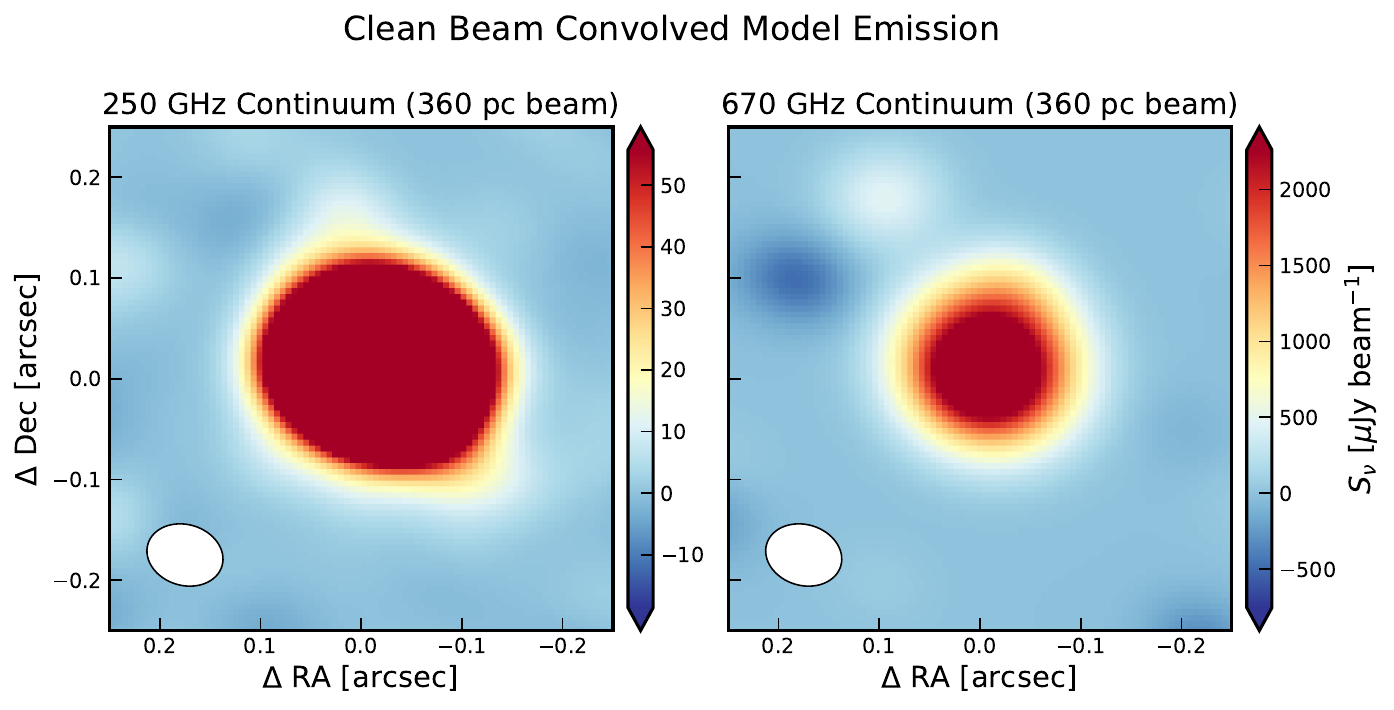}
    \includegraphics[width=0.9\linewidth]{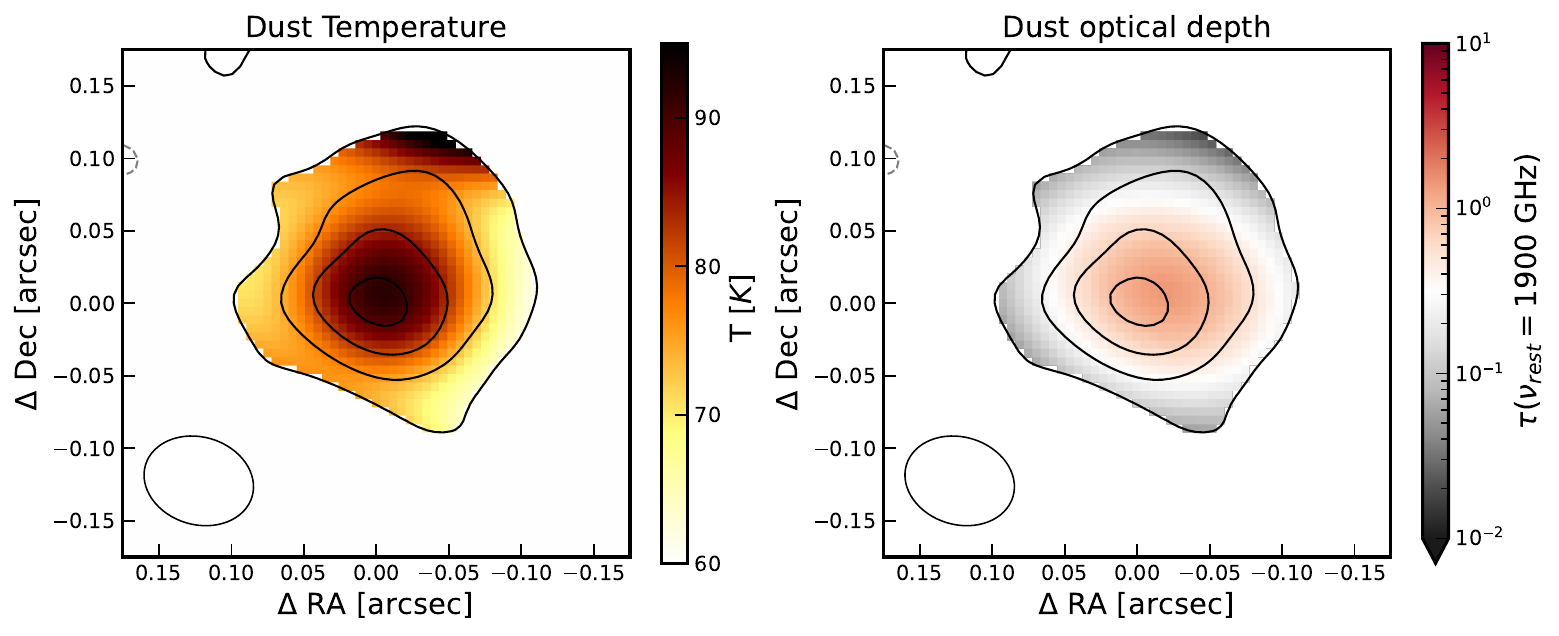}
    \caption{\textbf{Top row:} Model components (convolved by the clean beam) of the emission in the  continuum maps at $250$ and $670\ \rm{GHz}$. The color scales are the same as in Fig. \ref{fig:fig1}. \textbf{Bottom row: } Corresponding maps of the dust temperature (top left) and dust optical depth (top right). The color scales are the same as in \ref{fig:fig2_maps}. We observe a similar dust temperature and opacity gradient, confirming our results in Figure \ref{fig:fig2_maps} are not driven by the beam of the Band 6 or 9 observations.}
    \label{fig:maps_modelonly}
\end{figure*}

In this Appendix, we consider the impact of the dirty beam pattern on the dust temperature map inferred from the $250\ \rm{GHz}$ and $670\ \rm{GHz}$. Indeed, the final “cleaned" image produced by \emph{tclean} is the sum of the model emission convolved with the clean beam and the residuals below $2\sigma$, which still contain the dirty beam pattern. This leads to ill-defined units, a well-known problem which we address using residual-scaling for the integrated fluxes \citep[e.g.][]{Jorsater1995,Walter1999,Novak2019}. Here, we aim to check that the dirty beam imprinted in the residuals does not drive the temperature fluctuations reported in this paper.

To do, we subtract the residual maps from the cleaned image to obtain the model emission convolved with the clean beam at $250\ \rm{GHz}$ and $670\ \rm{GHz}$. Such maps are thus free of any dirty beam imprints from the residual. We then rescale the flux in these maps to match the aperture-integrated residual-scaled fluxes reported in \citet{Walter2022} and this work. We then proceed to fit each individual as done in the main body of this work. We show the model-only continuum emission and the inferred dust temperature and opacity maps in Figure \ref{fig:maps_modelonly}. We find that the dust temperature and opacity maps are consistent with what is observed when including the residuals, confirming that the gradients are real and not driven by the different dirty beams of the  $250\ \rm{GHz}$ and $670\ \rm{GHz}$ continuum observations.

\section{Impact of the optically-thick/thin assumption on the observed SED and inference of the dust properties}
\label{app:thin_thick_differences}
In this Appendix, we briefly highlight the different behaviour of optically-thin and -thick dust emission models with dust temperature and mass. This is of particular importance to the inference made when fitting the observed SED, especially with high frequency observations.

\begin{figure*}
    \centering
    \includegraphics[width=0.34\linewidth]{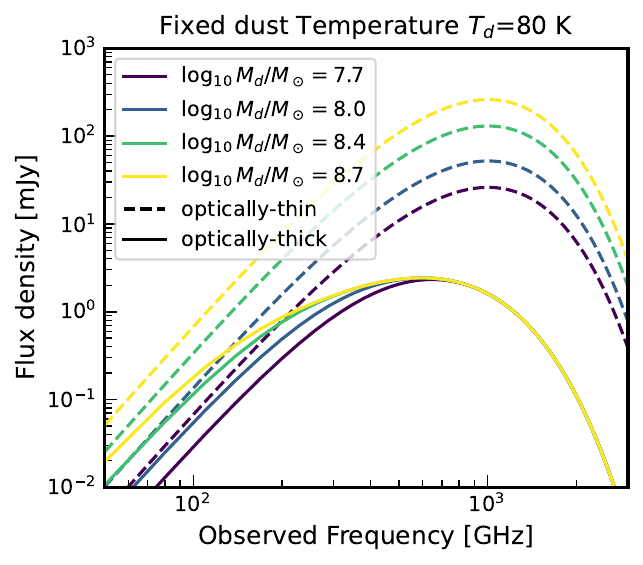}
    \includegraphics[width=0.62\linewidth]{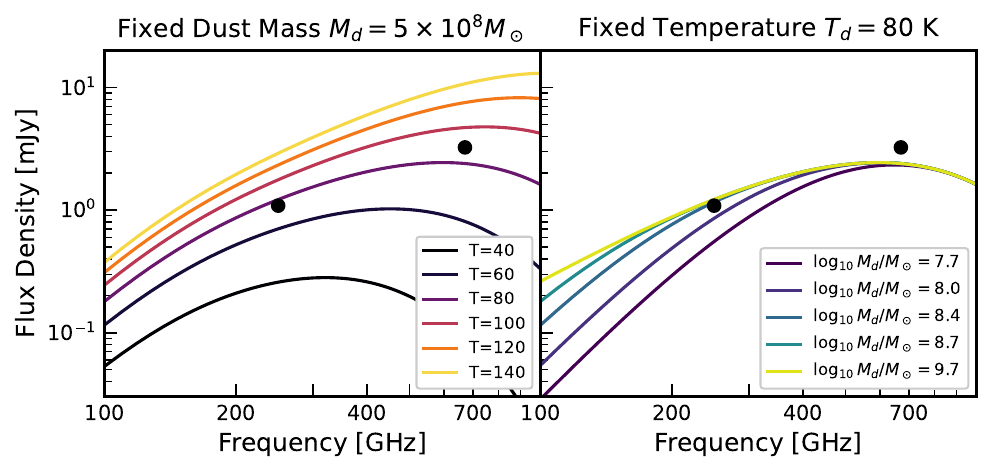}
    \caption{Illustrative plots showing the dust SED for various dust temperature and dust masses enclosed in an area equivalent to the central beam of our Band 9 observations. The black datapoints show the measured fluxes in the central beam (see Table \ref{tab:annuli_sed}). \textbf{Right:} Optically-thin versus optically-thick SED for various dust masses.  \textbf{Center and left:} Optically-thick dust SED for varying dust temperatures and dust masses. Under optically-thick conditions, higher dust masses do not translate to higher fluxes at higher frequencies, making Band 8-10 observations sensitive probes of the dust temperature. 
    }
    \label{fig:illustration_opt_thick}
\end{figure*}

In the case of optically-thin emission, adjusting the dust temperature and dust mass parameters can arbitrarily change the peak of the emission and the overall amplitude of the signal to match the data. In turn this requires a good sampling of the dust SED at various frequencies to constrain the dust properties. In the optically-thick case however, the emission is increasingly suppressed at high frequencies. The first consequence is that the dust SED peaks at lower frequencies than in the optically-thin case (see Fig. \ref{fig:illustration_opt_thick}, left), and the apparent slope is flatter.
Additionally, the amplitude of the SED at high frequencies plateaus when the optical depth (or equivalently the dust mass in a fixed area) increases further. Thus the high-frequency observations better constrain the dust temperature in the optically-thick case, even with a smaller number of constraints on FIR SED (see Fig. \ref{fig:illustration_opt_thick}). This is what enables us to determine relatively accurate resolved dust temperatures with only two high-resolution images in Band 6 and 9.

\section{Averaged Dust properties in radial annuli}
\label{app:dust_fit}
In this appendix, we detail the fitting procedure for the resolved and integrated FIR SED in J2348-3054, and present the resulting dust properties as a function of radius. The simultaneous fit to the resolved and integrated continuum observations is done as follows. We first assume that the AGN hot dust torus emission is unresolved at 360\ \rm{pc} resolution, and solely located at the peak of the Band 9 continuum observation. We then extract the fluxes in the central beam, and $3$ annuli spanning up to $0\farcs15$. For each annuli, we assume that the emission can be reproduced by an optically-thick dust emission model, and for the central resolution element we combine the dust emission with the hot dust torus emission. The total unresolved FIR SED is modeled with the sum of the four individual models detailed above. The total log-likelihood we aim to maximize is then simply the sum of the log-likelihoods for the model of each annuli compared to the resolved data, to which we add the likelihood of the global SED. As we assume errors are gaussian and i.i.d, the log likelihood is simply the sum of the $\chi^2$ for each annuli and the integrated SED. We fit the data at hand for each 1920 SKIRTOR models \citep{Stalevski2012,Stalevski2016} with nearly face-on inclination $i=10\textdegree$ (given that J2348-3054 is a luminous quasar but not a blazar), and select the model with the highest likelihood.

\begin{figure*}
    \centering
    \includegraphics[width=0.9\linewidth]{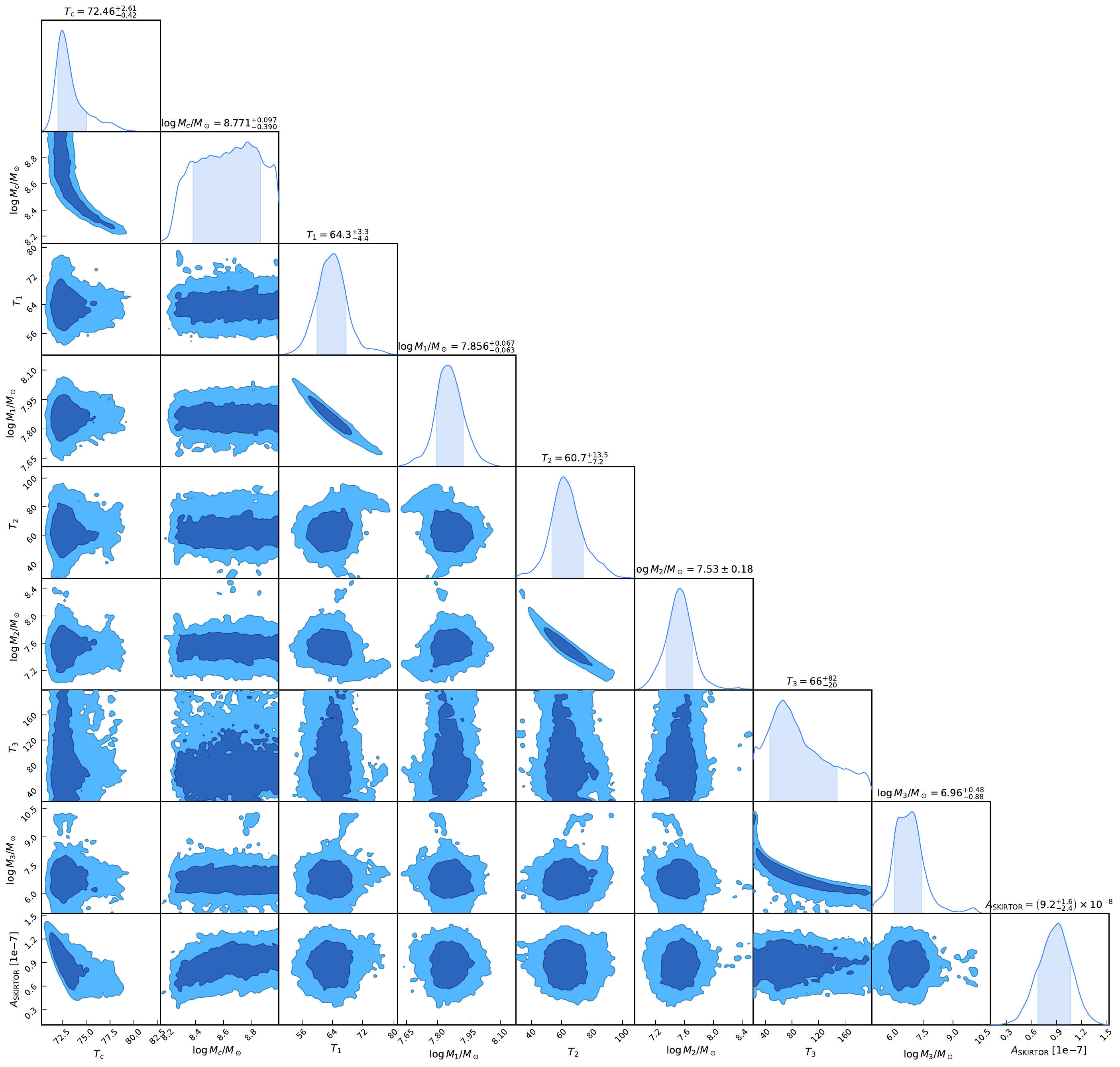}
    \caption{Parameter posterior distribution of the simultaneous fit to the resolved and integrated FIR observations. The index $c$ denotes the central beam, whilst the indices $i=1,2,3$ indicate the annuli at increasing radii. The last parameter is the renormalisation of the SKIRTOR model.}
    \label{fig:posterior}
\end{figure*}

We show the parameter posterior distribution in Fig. \ref{fig:posterior}, with most parameters being well constrained. The dust mass and temperature (and the hot dust torus contribution in the case of the central beam) are correlated for each annuli and the central beam, although correlation for parameters across radii are not significant. We then summarise the inferred dust temperatures, masses and FIR luminosities for each annuli in Table \ref{tab:annuli_sed}. For completeness we also present the same parameters inferred for independent fits to the each annuli where only the Band 6 and Band 9 data is considered, without the constraints from the total unresolved SED and without an AGN contribution.

\begin{table*}
\caption{Inferred average dust properties as a function of galactocentric radius in J2348--3054.} 
\centering
\def\arraystretch{1.5}% 
\begin{tabular}{cccccccccc}
\multicolumn{4}{}{} & \multicolumn{3}{c}{Independent annuli fits$^{a}$} & \multicolumn{3}{c}{Joint resolved+unresolved SED fit $^{b}$} \\
$r_{\rm{min}}$ & $r_{\rm{max}}$ & $S_\nu(250\  \rm{GHz})$  & $S_\nu(670\ \rm{GHz})$  & $M_d$   & $T_d$  & $L_{\rm{TIR}}$& $M_d$   & $T_d$  & $L_{\rm{TIR}}$  \\
$[\rm{pc}]$ & $[\rm{pc}]$  & [mJy] & [mJy] & [$10^8 M_\odot$] & [K] & [$10^{12}\ L_\odot$] & [$10^8 M_\odot$] & [K] & [$10^{12}\ L_\odot$] \\ \hline
0 & 216 & $1.086\pm0.006$ & $3.24\pm0.25$ & $1.17\pm 0.11$ & $88.2\pm2.3$  & $4.81^{-0.48}_{+0.47}$ & $5.9^{+1.4}_{-3.2}$ &$72.5^{+2.6}_{-0.4}$  & $2.23_{-0.02}^{+0.34}$  \\
216 & 253 & $0.596\pm0.010$ & $2.41\pm0.42$ &  $0.48\pm0.10$ & $63.3\pm5.7$ & $1.69^{-0.51}_{+0.67}$ & $0.72_{-0.10}^{+0.10}$ & $64.3^{+3.3}_{-4.4}$ & $2.73_{-0.61}^{+0.60}$ \\
253 & 380 & $0.504\pm0.014$ & $1.80\pm0.58$ &  $0.17\pm0.11$ & $61.0\pm17.9$ & $0.98^{-0.87}_{+0.52}$ & $0.34_{-0.11}^{+0.11}$  & $60.7^{+13.5}_{-7.2}$ & $1.94_{-0.64}^{+2.23}$ \\
380 & 506 & $0.147\pm0.018$ & $1.03\pm0.74$&  $0.06\pm0.13$ & $62.6\pm60.2$ & $0.52\pm0.52$ & $0.09_{-0.08}^{+0.18}$  & $66^{+82}_{-20}$ & $1.87_{-1.57}^{+15.75}$ \\ \hline
\multicolumn{2}{c}{Total} & & & $1.88\pm0.2$$^{a}$ & -- & $7.5\pm1.2$$^{a}$ & $5.7^{+3.3}_{-2.1}$ & -- & $8.78\pm0.10$
\end{tabular}
\tablefoot{The dust slope $\beta$ is fixed to $\beta=1.8$. See further Appendix \ref{app:dust_fit} for details on the simultaneous fitting of the resolved and integrated FIR SED. $^{a}$ Independent fits to the Band 6+9 resolved data, without the total SED constraint. Note that the sum of the luminosity and mass of the independent fits to each annuli are likely biased, as the sum of their best-fit SED is not constrained - and does not reproduced well - the total FIR SED observations. These results are only given here for completeness. $^{b}$ Joint fit to the unresolved and resolved data detailed further in appendix \ref{app:dust_fit}. The best-fit annuli SEDs and their sum (including the AGN torus contribution) are shown in Fig. \ref{fig:total_sed}. }
\label{tab:annuli_sed}
\end{table*}

\end{document}